\documentclass[12pt]{article}

\usepackage{amssymb,amsmath}

\usepackage{graphicx}
\DeclareGraphicsExtensions{.pdf,.png,.jpg}

 \catcode`\@=11 \@addtoreset{equation}{section}\catcode`\@=12
\newcommand{\R}{ {\mathbb R} }

\begin{document}

 \begin{center}

 \large \bf Stable  exponential cosmological  solutions with zero variation of $G$ 
 and three different Hubble-like parameters in the Einstein-Gauss-Bonnet  model  with a $\Lambda$-term 
  \end{center}

 \vspace{0.3truecm}

 \begin{center}

  K. K. Ernazarov$^{1}$ and V. D. Ivashchuk$^{1,2}$ 

\vspace{0.3truecm}

 \it $^{1}$Institute of Gravitation and Cosmology,
 RUDN University, 6 Miklukho-Maklaya ul.,
 Moscow 117198, Russia

 \it $^{2}$ Center for Gravitation and Fundamental Metrology,
 VNIIMS, 46 Ozyornaya ul., Moscow 119361, Russia

\end{center}

\begin{abstract}

We consider a $D$-dimensional  gravitational model with a Gauss-Bonnet term and the cosmological term $\Lambda$.
We restrict the metrics to diagonal cosmological ones and find for certain $\Lambda$ a class of solutions with  exponential time dependence of three scale factors, governed by three non-coinciding Hubble-like parameters $H >0$, $h_1$ and $h_2$, corresponding to factor spaces of dimensions $m > 2$, $k_1 > 1$ and $k_2 > 1$, respectively, with 
$k_1 \neq k_2$ and $D = 1 + m + k_1 + k_2$. Any of these solutions  describes an exponential expansion of  $3d$ subspace with Hubble parameter $H$ and  zero variation of the effective gravitational constant $G$. We  prove  the stability of these solutions in a class of cosmological solutions with diagonal  metrics.

\end{abstract}


\section{Introduction}

In this paper we consider a $D$-dimensional gravitational model
with Gauss-Bonnet term and cosmological term $\Lambda$.  
The so-called Gauss-Bonnet term appeared in string theory as a first order correction 
(in $\alpha'$) to the  effective action \cite{Zwiebach}-\cite{GW}.

 We note that at present the Einstein-Gauss-Bonnet (EGB) gravitational model and  its modifications, 
see   \cite{Ishihara}-\cite{ErIv-17} and refs. therein,  are intensively studied in  cosmology,
e.g. for possible explanation  of  accelerating  expansion of the Universe which follow from supernova (type Ia) observational data \cite{Riess,Perl,Kowalski}.

In ref. \cite{ErIv-17} we were dealing with the cosmological solutions with diagonal metrics  
governed by $n >3$ scale factors
depending upon one variable, which is the synchronous time variable. We have restricted ourselves by the
solutions with exponential dependence of scale factors and have presented 
a class of such solutions with  two scale factors, governed by two Hubble-like parameters $H >0$ and $h < 0$, which correspond to factor spaces of dimensions $m > 3$ and $l > 1$, respectively, with $D = 1 + m + l$ and $(m,l) \neq  (6,6), (7,4), (9,3)$.
Any of  these solutions  describes   an exponential expansion  of $3d$ subspace with Hubble parameters $H > 0$ \cite{Ade}  and has a  constant volume factor  of  $(m - 3 + l)$-dimensional internal space,
which implies zero variation of the effective gravitational constant $G$ either in a Jordan or  an Einstein frame \cite{RZ-98,I-96}; see also  \cite{BIM,Mel,IvMel-14} and refs. therein. These solutions satisfy the most severe restrictions on  variation of $G$ \cite{Pitjeva}. We have  studied  the stability of these solutions in a class of cosmological solutions with diagonal metrics by using results of  refs. \cite{ErIvKob-16,Ivas-16} (see also approach of ref. \cite{Pavl-15}) and have shown  that all solutions, presented in ref. \cite{ErIv-17}, are stable. 
It should be noted that  two  special solutions  for $D = 22, 28$ and $\Lambda = 0$ were found earlier in ref.  \cite{IvKob}; in ref. \cite{ErIvKob-16} it was proved that these solutions are stable.
Another set of six stable  exponential solutions, five in dimensions $D = 7, 8, 9, 13$ and two  for $D = 14$, were considered earlier in \cite{Ivas-16-2}. 

In this paper we extend the results of ref. \cite{ErIv-17} to the case of solutions with 
three non-coinciding Hubble-like parameters. The structure of the paper is as follows. In Section 2 we present a setup. A class of exact cosmological solutions with   diagonal metrics is found  for certain $\Lambda$ in Section 3.
Any of these solutions  describes an exponential expansion of  3-dimensional subspace with Hubble parameter $H$ and  zero variation of the effective gravitational constant $G$. In Section 4 we  prove  the stability of the solutions in a class of cosmological solutions with diagonal  metrics. Certain examples are presented in Section 5.

\section{The cosmological model}

The action of the model reads
\begin{equation}
  S =  \int_{M} d^{D}z \sqrt{|g|} \{ \alpha_1 (R[g] - 2 \Lambda) +
              \alpha_2 {\cal L}_2[g] \},
 \label{2.0}
\end{equation}
where $g = g_{MN} dz^{M} \otimes dz^{N}$ is the metric defined on
the manifold $M$, ${\dim M} = D$, $|g| = |\det (g_{MN})|$, $\Lambda$ is
the cosmological term, $R[g]$ is scalar curvature,
$${\cal L}_2[g] = R_{MNPQ} R^{MNPQ} - 4 R_{MN} R^{MN} +R^2$$
is the standard Gauss-Bonnet term and  $\alpha_1$, $\alpha_2$ are
nonzero constants.

We consider the manifold
\begin{equation}
   M = \R  \times   M_1 \times \ldots \times M_n 
   \label{2.1}
\end{equation}
with the metric
\begin{equation}
   g= - d t \otimes d t  +
      \sum_{i=1}^{n} B_i e^{2v^i t} dy^i \otimes dy^i,
  \label{2.2}
\end{equation}
where   $B_i > 0$ are arbitrary constants, $i = 1, \dots, n$, and
$M_1, \dots,  M_n$  are one-dimensional manifolds (either $\R$ or $S^1$)
and $n > 3$.

The equations of motion for the action (\ref{2.0}) 
give us the set of  polynomial equations \cite{ErIvKob-16}
\begin{eqnarray}
  E = G_{ij} v^i v^j + 2 \Lambda
  - \alpha   G_{ijkl} v^i v^j v^k v^l = 0,  \label{2.3} \\
   Y_i =  \left[ 2   G_{ij} v^j
    - \frac{4}{3} \alpha  G_{ijkl}  v^j v^k v^l \right] \sum_{i=1}^n v^i 
    - \frac{2}{3}   G_{ij} v^i v^j  +  \frac{8}{3} \Lambda = 0,
   \label{2.4}
\end{eqnarray}
$i = 1,\ldots, n$, where  $\alpha = \alpha_2/\alpha_1$. Here
\begin{equation}
G_{ij} = \delta_{ij} -1, \qquad   G_{ijkl}  = G_{ij} G_{ik} G_{il} G_{jk} G_{jl} G_{kl}
\label{2.4G}
\end{equation}
are, respectively, the components of two  metrics on  $\R^{n}$ \cite{Iv-09,Iv-10}. 
The first one is a 2-metric and the second one is a Finslerian 4-metric.
For $n > 3$ we get a set of forth-order polynomial  equations.

We note that for $\Lambda =0$ and $n > 3$ the set of equations (\ref{2.3}) 
and (\ref{2.4}) has an isotropic solution $v^1 = \cdots = v^n = H$ only 
if $\alpha  < 0$ \cite{Iv-09,Iv-10}.
This solution was generalized in \cite{ChPavTop} to the case $\Lambda \neq 0$.

It was shown in \cite{Iv-09,Iv-10} that there are no more than
three different  numbers among  $v^1,\dots ,v^n$ when $\Lambda =0$. This is valid also
for  $\Lambda \neq 0$ if $\sum_{i = 1}^{n} v^i \neq 0$  \cite{Ivas-16}.

\section{Solutions with constant $G$}

In this section we present a class of solutions to the set of equations (\ref{2.3}), 
(\ref{2.4}) of the following form:
\begin{equation}
  \label{3.1}
   v =(\underbrace{H,H,H}_{``our'' \ space},\underbrace{\overbrace{H, \ldots, H}^{m-3}, 
   \overbrace{h_1, \ldots, h_1}^{k_1}, \overbrace{h_2, \ldots, h_2}^{k_2}}_{internal \ space}).
\end{equation}
where $H$ is the Hubble-like parameter corresponding  
to an $m$-dimensional factor space with $m > 2$,  $h_1$ is the Hubble-like parameter 
corresponding to an $k_1$-dimensional factor space with $k_1 > 1$ and $h_2$ ($h_2 \neq h_1$) is the Hubble-like parameter 
corresponding to an $k_2$-dimensional factor space with $k_2 > 1$. We split the $m$-dimensional  
factor space into the  product of two subspaces of dimensions $3$ and $m-3$, respectively. 
The first one is identified with ``our'' $3d$ space while the second one is considered as 
a subspace of $(m-3 + k_1 + k_2)$-dimensional internal space.
 
We put 
\begin{equation}
  \label{3.2a}
   H > 0 
\end{equation}
for a description of an accelerated expansion of a
$3$-dimensional subspace (which may describe our Universe) and also put
\begin{equation}
  \label{3.2}
(m-3) H + k_1 h_1 +  k_2 h_2 = 0
\end{equation}
for a  description of a zero variation of the effective gravitational constant $G$.

We remind (the reader) that  the effective gravitational constant $G = G_{eff}$ in the Brans-Dicke-Jordan (or simply Jordan) frame \cite{RZ-98} (see also \cite{I-96})
is proportional to the inverse volume scale factor
of the internal space; see \cite{BIM,Mel,IvMel-14} and references therein.

Due to  (\ref{3.1}) ``our'' 3d space expands isotropically with 
Hubble parameter $H$, while the $(m -3)$-dimensional part of 
the internal space expands isotropically with the same Hubble parameter $H$ too. 
Here, like in ref. \cite{ErIv-17}, we consider  for 
cosmological applications (in our epoch) the internal space to be compact one, i.e. we  put in (\ref{2.1}) 
$M_4 = \cdots = M_n = S^1$. We  put the internal scale factors corresponding to present time $t_0$ :  $a_j (t_0) = B_j^{1/2} \exp(v^j t_0) $, $j =4, \ldots, n$, (see (\ref{2.2})) to be small enough in comparison with the scale factor of ``our'' space for $t = t_0$: $ a (t_0)  = B^{1/2} \exp(H t_0) $, where  $B_1 = B_2 = B_3 = B$.                    

According to the ansatz (\ref{3.1}),  the $m$-dimensional factor space is expanding with the Hubble parameter $H >0$, while the $k_i$-dimensional factor space  is contracting with the Hubble-like  parameter $h_i < 0$, where $i$ is 
either $1$ or $2$.

Now we consider the ansatz (\ref{3.1}) with three Hubble parameters $H$, $h_1$ and $h_2$ 
which obey the following restrictions:
   \begin{equation}
   S_1 = m H + k_1 h_1 + k_2 h_2 \neq 0, \quad  H \neq h_1, \quad  H \neq h_2, \quad  h_1 \neq h_2.
   \label{3.3}
   \end{equation}
The first inequality in (\ref{3.3}) is valid since $S_1 = 3H > 0$  
due to (\ref{3.2a}) and (\ref{3.2}).

   In this  case the set of $n+1$ eqs. (\ref{2.3}), (\ref{2.4})
   is equivalent to the set of three equations
    \begin{equation}
     E =0, \qquad   Y_H = 0, \qquad  Y_{h_1} = 0,  \qquad  Y_{h_2} = 0,
       \label{3.4}
    \end{equation}
   where  
   \begin{equation}
   Y_{H} = Y_{\mu}, \quad Y_{h_1} = Y_{\alpha}, \quad   Y_{h_2} = Y_{a}, 
        \label{3.4a}
    \end{equation}
   for all $\mu = 1, \dots, m$; $\alpha = m + 1, \dots, m + k_1$ and $a = m + k_1 + 1, \dots, n$.
   These  relations follow from the definition of $Y_{i}$ in (\ref{2.4}) and the identities 
   \cite{Iv-09,Iv-10}
     \begin{eqnarray}
    v_i = G_{ij}v^j = v^i - S_1,
     \label{3.4b}   \\
    A_i =  G_{ijkl} v^j v^k v^l
          = S_1^3  + 2 S_3 -3 S_1 S_2
    \nonumber \\
      +  3 (S_2  - S_1^2)  v^i
      +  6 S_1 (v^i)^2 - 6(v^i)^3,
     \label{3.4c}
     \end{eqnarray}
   $i = 1,\ldots, n$, where here and in what follows 
    \begin{equation}
   S_k = \sum_{i =1}^n (v^i)^k.
   \label{3.4d}
    \end{equation}     

   Due to  (\ref{2.4}), (\ref{3.4b}), (\ref{3.4c}) 
    we obtain
    \begin{equation}
       Y_{h_i} - Y_{h_j} = (h_i - h_j) S_1 [2 + 4 \alpha Q_{h_i,h_j}],
           \label{3.4e}
    \end{equation}
     where
     \begin{equation}
     Q_{h_i h_j} =  S_1^2 - S_2 - 2 S_1 (h_i + h_j) + 2 (h_i^2 + h_i h_j + h_j^2),
                \label{3.5}
     \end{equation}
     $i \neq j$; $i,j =0,1,2$ and $h_0 = H$. 
 Relations   (\ref{3.3}), (\ref{3.4}) and (\ref{3.4e}) imply 
   \begin{equation}
     Q_{h_i h_j} =  - \frac{1}{2 \alpha},
   \label{3.5a}
   \end{equation}  
     $i \neq j$ and $i,j =0,1,2$.

 Due to $S_1 = m H + k_1 h_1 + k_2 h_2 \neq 0$ the set of eqs.  (\ref{3.4})
 is equivalent to the following set of equations
 \begin{eqnarray}
  E =0, \quad   Y_H - Y_{h_1} = 0,  \quad   Y_{h_1} - Y_{h_2} = 0, 
  \nonumber    \\
  \quad  m H Y_H + k_1 h_1 Y_{h_1} + k_2 h_2 Y_{h_2} = 0.
       \label{3.5b}
  \end{eqnarray}
  The last relation in (\ref{3.5b}) may be omitted since $E = 0$ implies 
  $Y_{i}h^i = m H Y_H + k_1 h_1 Y_{h_1} + k_2 h_2 Y_{h_2} = 0$ \cite{Ivas-16}. Using this fact 
  and relations (\ref{3.3}) and  (\ref{3.4e}) we reduce the system  (\ref{3.5b})
  to the following one
  \begin{equation}
    E =0, \quad   Q_{H h_1} =  - \frac{1}{2 \alpha},  \quad   Q_{h_1 h_2} =  - \frac{1}{2 \alpha}. 
     \label{3.5c}
    \end{equation}
  Using the identity
  \begin{equation}
     Q_{H h_1}  - Q_{h_1 h_2} = (H - h_2) (-S_1 + H + h_1 + h_2), 
       \label{3.5d}
 \end{equation} 
  we reduce the set of equations  (\ref{3.5c}) to the equivalent set 
  \begin{equation}
      E =0, \quad   Q =  - \frac{1}{2 \alpha},  \quad     H + h_1 + h_2 - S_1 = 0. 
       \label{3.5cc}
      \end{equation}
  Here  we put $Q = Q_{h_1 h_2}$ though other choices, $Q = Q_{H h_1}$ or $Q = Q_{H h_2}$ 
  give us equivalent sets of equations. Thus the set of $(n + 1)$ polynomial equations
  (\ref{2.3}), (\ref{2.4}) under ansatz  (\ref{3.1}) and restrictions (\ref{3.3}) imposed
  is reduced to a set  (\ref{3.5cc}) of three polynomial equations 
  (of fourth, second and first orders). This reduction is a special case of the more general
  prescription from ref. \cite{ChPavTop1}.
 
Using the condition (\ref{3.2}) of zero variation of  $G$ and the linear equation from (\ref{3.5cc})
we obtain for $k_1 \neq k_2$ 
 \begin{equation}
   h_1 = \frac{m + 2 k_2 - 3}{k_2 - k_1} H, \qquad  h_2 = \frac{m + 2 k_1 - 3}{k_1 - k_2} H. 
    \label{3.5dd}
 \end{equation}
 For $k_1 = k_2$ we get $H=0$, which is not appropriate for our consideration. 

The substitution of (\ref{3.5dd}) into relation $Q_{h_1 h_2} =  - \frac{1}{2 \alpha}$
gives us the following relation
 \begin{equation}
  \frac{P}{(k_2 - k_1)^2} H^2 = - \frac{1}{2 \alpha}, 
    \label{3.5e}
 \end{equation}
for $k_1 \neq k_2$, where 
\begin{eqnarray}
P  =  P(m,k_1,k_2)  =- (m + k_1 + k_2 -3)(m (k_1 + k_2 -2) +
\nonumber \\
 k_1 ( 2 k_2 -5 ) + k_2 ( 2 k_1 -5 ) + 6) \neq 0, 
 \label{3.7} 
\end{eqnarray}
which implies
\begin{equation}
H   =   |k_1 - k_2| (- 2 \alpha P)^{-1/2}, \qquad \alpha P < 0. 
\label{3.8}
\end{equation}

It may be readily verified that 
\begin{equation}
P  =  P(m,k_1,k_2) < 0  
\label{3.8p}
\end{equation}
for all $m > 2$,  $k_1 > 1$, $k_2 > 1$, $k_1 \neq k_2$
and hence our solutions take place for $\alpha > 0$.

The substitution of (\ref{3.5dd}) into (\ref{3.4})
gives us 
\begin{equation}     
2 \Lambda = - F_1 H^2 - F_2 H^4
 \label{3.9} 
\end{equation}
where
\begin{eqnarray}
 F_{1} =  \frac{1}{(k_{2}-k_{1})^2} [(k_{1}+k_{2})m^2+(k_{1}^2 +
 6k_{1}k_{2}+ k_{2}^2 - 6k_{1} -6k_{2}) m 
 \nonumber \\
 - 9(k_{1}^2+k_{2}^2 - k_{1} - k_{2})+2(2k_{1}+2k_{2}-3)k_{1}k_{2}]  \quad \quad \quad \quad
 \label{3.10} 
\end{eqnarray}
and
\begin{eqnarray}
  F_{2}= & {} -  \, \frac{3\alpha (m-3+k_{1}+k_{2})}{(k_{2}-k_{1})^4} [(k_{1}+k_{2})(k_{1}+k_{2}-2)m^3 
   \nonumber \\
   & + \, (k_{1}+k_{2})(k_{1}^2+k_{2}^2 +10k_{1}k_{2}-15(k_{1}+k_{2})+18)m^2 \nonumber \\
   & - \, (12(k_{1}^3+k_{2}^3)-63(k_{1}^2+k_{2}^2)+54(k_{1}+k_{2}) \nonumber \\
   &- \, 2(4(k_{1}^2+k_{2}^2)-42(k_{1}+k_{2}+16k_{1}k_{2}+63)k_{1}k_{2}))m \nonumber \\
   &+ \, 27(k_{1}^3+k_{2}^3)-81(k_{1}^2+k_{2}^2) +54(k_{1} +k_{2})  \nonumber \\
   & - \, (40(k_{1}^2+k_{2}^2)-16(k_{1}+k_{2}-6)k_{1}k_{2}+162 \nonumber \\
   & - \, 153(k_{1}+k_{2}))k_{1}k_{2}].     \label{3.11}                  
\end{eqnarray}

Using relations (\ref{3.8}), (\ref{3.9}), (\ref{3.10}), (\ref{3.11}) we obtain
\begin{eqnarray}
   & {} \Lambda =   \Lambda(m,k_1,k_2) = \frac{1}{8 \alpha P^2} (m + k_1 + k_2 -3)
   \nonumber \\ 
   & \times \, [(k_1 + k_2)(k_1 + k_2 - 2)m^3  
   \nonumber \\
   &+ \, (k_1^3 + k_2^3 + 11 (k_1^2 k_2 + k_1 k_2^2) - 19 (k_1^2 + k_2^2) 
   \nonumber \\
   &- \, 22 k_1 k_2 + 18 (k_1 + k_2)) m^2
   \nonumber \\
   & - \, (8(k_1^3 + k_2^3)  - 63 (k_1 + k_2)^2 - 8 k_1^2 (k_1 - 11) k_2
    \nonumber \\
   & - \, 8 k_2^2 (k_2 - 11) k_1  - 32 k_1^2 k_2^2 + 54 (k_1 + k_2)) m 
   \nonumber \\ 
   & - \, ( 9 (k_1^3 + k_2^3) + 45 (k_1^2 + k_2^2) - 54 (k_1 + k_2)
     \nonumber \\
   & + \, 8  (k_1^2 + k_2^2) k_1 k_2  \nonumber \\ 
   & - \, 16 (k_1 + k_2  -10) k_1^2 k_2^2 - 9 (21 k_1 + 21 k_2  - 26) k_1 k_2 ) ], 
    \label{3.12}
\end{eqnarray}
where  $P  =  P(m,k_1,k_2)$ is defined in  (\ref{3.7}).
 
 The function $\Lambda(m,k_1,k_2)$ in (\ref{3.12}) is symmetric with respect to $k_1$ and $k_2$, i.e. 
 \begin{equation}     
 \Lambda(m,k_1,k_2) = \Lambda(m,k_2,k_1).
  \label{3.13} 
 \end{equation}
For  $k_2 =0$ we get a function $\Lambda(m,k_1,0) = \Lambda(m,k_1)$, where $\Lambda(m,k_1)$
was obtained in ref. \cite{ErIv-17} for  the case  of two different Hubble-like parameters.

It may be readily verified that for $k_1(k) = n_1 k +q_1$ and  $k_2(k) = n_2 k +q_2$, 
where $k$,  $n_1 > 0$, $q_1$, $n_2 > 0$, $q_2$ are integer numbers, we get  
 \begin{equation}     
  \Lambda(m,k_1(k),k_2(k)) \to  \frac{1}{8\alpha},
  \label{3.14} 
 \end{equation}
as $k \to + \infty$ for any fixed $m \geq 3$. We note that the limit (\ref{3.14}) is positive 
and  does not depend upon $m$.
For fixed  integer $m > 2$ and $k_2 \geq 1$ we are led to the following limit
  \begin{eqnarray}     
   \Lambda(m,k_1,k_2) \to &  \frac{1}{8\alpha (m + 4 k_2 -5)^2} [ m^2 - 8(1 - k_2) m 
   \nonumber  \\ 
      &- \, 9 - 8 k_2 + 16 k_2^2 ]  = \Lambda(m,\infty,k_2),
   \label{3.15} 
  \end{eqnarray}
 as $k_1 \to + \infty$ and analogous relation (due to (\ref{3.13}))  for fixed  $m > 2$,  $k_1 \geq 1$
 and $k_2 \to + \infty$. It can be easily verified that for these values 
 of $m$,  $k_1$ we get: $\Lambda(m,\infty,k_2) > 0$.
 
 Relation  (\ref{3.14}) and (\ref{3.15}) may be used in a context of $(1/D)$-expansion 
 for large $D$ in the model under consideration,
 see  \cite{CGPT} and refs. therein.

\section{The proof of stability}

Here, as in \cite{ErIv-17}, we have due to (\ref{3.2})
\begin{equation}
  K = K(v) = \sum_{i = 1}^{n} v^i = 3H >0.
  \label{4.1}
\end{equation}

Let us put  the  restriction 
\begin{equation}
  \det (L_{ij}(v)) \neq 0
  \label{4.2}
\end{equation}
on the matrix 
\begin{equation}
L =(L_{ij}(v)) = (2 G_{ij} - 4 \alpha G_{ijks} v^k v^s).
   \label{4.1b}
 \end{equation}

We remind that for general cosmological setup with the metric 
\begin{equation}
 g= - dt \otimes dt + \sum_{i=1}^{n} e^{2\beta^i(t)}  dy^i \otimes dy^i,
 \label{4.3}
\end{equation}
we have  the  set of  equations \cite{ErIvKob-16} 
\begin{eqnarray}
     E = G_{ij} h^i h^j + 2 \Lambda  - \alpha G_{ijkl} h^i h^j h^k h^l = 0,
         \label{4.3.1} \\
         Y_i =  \frac{d L_i}{dt}  +  (\sum_{j=1}^n h^j) L_i -
                 \frac{2}{3} (G_{sj} h^s h^j -  4 \Lambda) = 0,
                     \label{4.3.2a}
          \end{eqnarray}
where $h^i = \dot{\beta}^i$,           
 \begin{equation}
  L_i = L_i(h) = 2  G_{ij} h^j
       - \frac{4}{3} \alpha  G_{ijkl}  h^j h^k h^l  
       \label{4.3.3},
 \end{equation}
 $i = 1,\ldots, n$.

Due to results of ref. \cite{Ivas-16}  a fixed point solution
$(h^i(t)) = (v^i)$ ($i = 1, \dots, n$; $n >3$) to eqs. (\ref{4.3.1}), (\ref{4.3.2a})
obeying restrictions (\ref{4.1}), (\ref{4.2}) is  stable under perturbations
\begin{equation}
 h^i(t) = v^i +  \delta h^i(t), 
\label{4.3h}
\end{equation}
 $i = 1,\ldots, n$,  as $t \to + \infty$.

In order to prove the stability of solutions  we should prove relation (\ref{4.2}).
First, we show that  for  the vector $v$ from  (\ref{3.1}), obeying
 relations (\ref{3.3}) the matrix $L$ has a block-diagonal form
\begin{equation}
 (L_{ij}) = {\rm diag}(L_{\mu \nu}, L_{\alpha \beta}, L_{a b} ),
 \label{4.5}
\end{equation}
where  here and in what follows: $\mu, \nu = 1, \dots, m$; $\alpha, \beta = m + 1, \dots, m + k_1$ 
and $a, b = m + k_1 + 1, \dots, n$.

Indeed, denoting $S_{ij} = G_{ijkl} v^k v^l$ we get from  (\ref{3.4c})
\begin{eqnarray}
S_{ij} = \frac{1}{3} \frac{\partial}{\partial v^j} (G_{iskl} v^s v^k v^l)  \nonumber \\
 = S_1^2 - S_2 + 2 (v^i)^2 + 2 (v^j)^2 + 2 v^i v^j -  2 S_1 (v^i + v^j)  \nonumber \\
  +  \delta_{ij} (S_2  - S_1^2   +  4 S_1 v^i - 6(v^i)^2).
 \label{4.5S}     
\end{eqnarray}    
Here we use the notation  $S_k = \sum_{i =1}^{n} (v^i)^k$ and the identity 
$ \frac{\partial}{\partial v^j}  S_k = k (v^j)^{k - 1}$.
It follows from (\ref{3.5}) and  (\ref{4.5S}) that
\begin{eqnarray}
S_{H h_1} \equiv  S_{\mu \alpha} = S_{ \alpha \mu} = Q_{H h_1},   
   \label{4.5SQa}   \\
S_{H h_2} \equiv S_{\mu a} = S_{a \mu} = Q_{H h_2}, 
   \label{4.5SQb}   \\
S_{h_1 h_2} \equiv S_{\alpha a} = S_{a \alpha} = Q_{h_1 h_2}
  \label{4.5SQc}
\end{eqnarray} 
and hence $L_{\mu \alpha} = L_{ \alpha \mu} = 0$,  
 $L_{\mu a} = L_{a \mu} = 0$ and  $L_{\alpha a} = L_{a \alpha} = 0$ 
 due to eqs.  (\ref{3.5a}). Thus, the matrix $(L_{ij})$ is a block-diagonal
 one.

For other three blocks we have
\begin{eqnarray}
  L_{\mu \nu} =  G_{\mu \nu} (2 + 4 \alpha S_{HH}),
 \label{4.6HH}     \\
  L_{\alpha \beta} = G_{\alpha \beta} (2 + 4 \alpha S_{h_1 h_1}), 
  \label{4.6h1h1} \\
   L_{a b} = G_{a b} (2 + 4 \alpha S_{h_2 h_2}),
   \label{4.6h2h2}
  \end{eqnarray}
where 
\begin{equation}
  S_{h_i h_i} =  S_1^2 - S_2 + 6 h_i^2 - 4 S_1 h_i,
  \label{4.7} 
\end{equation}
$i = 0,1,2$ and  $h_0 = H$. Here we denote: $S_{HH} = S_{\mu \nu}$, $\mu \neq \nu$;
$S_{h_1 h_1} = S_{\alpha \beta}$, $\alpha \neq \beta$ and 
$S_{h_1 h_1} = S_{a b}$, $a \neq b$. 
 
Due to relations  (\ref{4.5}), (\ref{4.6HH}), (\ref{4.6h1h1}), (\ref{4.6h2h2})
the matrix (\ref{4.5}) is invertible if and only if $m > 1$,  $k_1 > 1$, $k_2 > 1$ and
 \begin{equation}
   S_{h_i h_i} \neq - \frac{1}{2 \alpha},  
  \label{4.8}  
  \end{equation}
 $i = 0,1,2$.

Now, we prove that inequalities (\ref{4.8}) are satisfied for the solutions 
under consideration. Let us suppose that   (\ref{4.8}) is not satisfied for some 
$i_0 \in \{0,1,2 \}$, i.e. 
\begin{equation}
  S_{h_{i_0} h_{i_0}} =  S_1^2 - S_2 + 6 h_{i_0}^2 - 4 S_1 h_{i_0} = - \frac{1}{2 \alpha}.  
 \label{4.9}  
 \end{equation}
Let  $i_1 \in \{0,1,2 \}$ and $i_1 \neq i_0$.  Then using relations (\ref{3.5})
   and (\ref{3.5a})  we get
\begin{equation}
   Q_{h_{i_0} h_{i_1}}  - S_{h_{i_0} h_{i_0}} = 2 (h_{i_1} - h_{i_0}) ( 2 h_{i_0} + h_{i_1} - S_1) = 0,  
 \label{4.10}  
 \end{equation}    
 which implies 
 \begin{equation}
    2 h_{i_0} + h_{i_1} - S_1 = 0.  
  \label{4.11}  
  \end{equation} 
   But due to (\ref{3.5cc}) 
  \begin{equation}
       h_{i_0} + h_{i_1} + h_{i_2} - S_1 = 0,  
    \label{4.12}
  \end{equation}  
 where $i_2 \in \{0,1,2 \}$ and $i_2 \neq i_0$, $i_2 \neq i_1$. Subtracting  (\ref{4.12})
 from  (\ref{4.11}) we obtain  $h_{i_0} - h_{i_2} = 0$, i.e. $h_{i_0} = h_{i_2}$.
 But due to restrictions (\ref{3.3}) we have $h_{i_0} \neq h_{i_2}$. We are led 
 to a contradiction, which proves the inequalities (\ref{4.8}) and hence 
  the matrix $L$ from (\ref{4.5}) is invertible ($m > 2$,  $k_1 > 1$, $k_2 > 1$), 
  i.e. relation (\ref{4.2}) is obeyed.  Thus,  the solutions under consideration  are stable.  

\section{Examples}

Here we present several examples of stable solutions under consideration.

\subsection{The case m =3}

Let us consider the case $m =3$. From (\ref{3.12}) we get

 \begin{eqnarray}
   \Lambda =- \frac{1}{4 \alpha}\frac{1}{(k_{1}- 2k_{1}k_{2}+ k_{2})^2 (k_{1}+k_{1})} \times \nonumber\\
\times (3(k_{1}^3+k_{2}^3)-(2(k_{1}^2+k_{2}^2)+(k_{1}+k_{2})(3+2k_{1}k_{2})-8k_{1}k_{2})k_{1}k_{2})
\label{5.1}
\end{eqnarray}
 
 For $(m,k_1,k_2) = (3,3,2)$ we have $P = -70$,
 \begin{equation}
 \Lambda =\frac{213}{980 \alpha}
 \label{5.2}
 \end{equation}
and 
 \begin{equation}
  H = \frac{1}{\sqrt{140 \alpha}}, \qquad h_{1}=-4H, \qquad  h_{2}=6H.
  \label{5.3}                                                                                                        \end{equation}

Now we put $(m,k_1,k_2) = (3,4,2)$. We obtain $P = -120$ 
 \begin{equation}
 \Lambda = \frac{21}{100\alpha},
 \label{5.4}
 \end{equation}
 and
 \begin{equation}
  H =\frac{1}{2\sqrt{15 \alpha}}, \qquad  h_{1} = -2H, \qquad  h_{2} = 4H.
  \label{5.5}
\end{equation}
According to our analysis from the previous section both solutions are stable.

\subsection{Examples for $m =4$ and $m= 5$}

Now we present other examples of stable solutions for $m =4$ and $m= 5$.

First we put $(m,k_1,k_2) = (4,3,2)$. We find $P = -102$ and 
 \begin{equation}
 \Lambda = \frac{123}{578\alpha}.
 \label{5.6}
\end{equation}
In this case we obtain 
 \begin{equation}
  H = \frac{1}{\sqrt{204 \alpha}}, \qquad  h_{1} = -5H, \qquad    h_{2}=7H.
\label{5.7}
\end{equation}

Now we enlarge the value of $m$ by putting  $(m,k_1,k_2) = (5,3,2)$. We find $P = -140$,
\begin{equation}
 \Lambda=\frac{589}{2800\alpha}
 \label{5.8}
\end{equation}
and
 \begin{equation}
  H=\frac{1}{\sqrt{280 \alpha}}, \qquad  h_{1}= - 6H, \qquad   h_{2} = 8H.
 \label{5.9}
\end{equation}

We note that in all examples above $\Lambda >0$. 

\section{Conclusions}

We have considered the  $D$-dimensional  Einstein-Gauss-Bonnet (EGB) model
with the $\Lambda$-term and two constants $\alpha_1$ and $\alpha_2$.  By using the  ansatz with diagonal  cosmological  metrics, we have found, for certain  $\Lambda = \Lambda(m,k_1.k_2)$ and  
$\alpha = \alpha_2 / \alpha_1 < 0$, a class of solutions with  exponential time dependence of three scale factors, governed by three different Hubble-like parameters $H >0$,  $h_1$ and $h_2$, corresponding to submanifolds of 
dimensions $m > 2$, $k_1 > 1$, $k_2 > 1$, respectively, with $ k_1 \neq k_2$ and $D = 1 + m + k_1 + k_2 $.
Here $m > 2$ is the dimension  of the expanding subspace.

Any of these solutions describes an exponential expansion of ``our'' 3-dimensional subspace with
the Hubble parameter $H > 0$ and anisotropic behaviour of $(m-3+ k_1 + k_2)$-dimensional internal space:
expanding in $(m-3)$ dimensions (with Hubble-like parameter $H$) and either 
contracting  in $k_1$ dimensions (with Hubble-like parameter $h_1$) and expanding  
in $k_2$ dimensions (with Hubble-like parameter $h_2$) for $k_1 > k_2$ or 
expanding in $k_1$ dimensions  and contracting in $k_2$ dimensions for $k_1 < k_2$.
Each solution   has a constant volume factor of internal space and hence it describes
zero variation of the  effective gravitational constant $G$.  By using results of ref. \cite{Ivas-16}
we  have proved that all these solutions  are stable as  $t \to + \infty$. We have presented 
several examples of stable solutions for $m = 3,4,5$.

 {\bf Acknowledgments}

This paper was funded by the Ministry of Education and Science of the Russian Federation
in the Program to increase the competitiveness of Peoples� Friendship University
(RUDN University) among the world's leading research and education centers in the 2016-2020
and  by the  Russian Foundation for Basic Research,  grant  Nr. 16-02-00602.

\end{document}